\documentclass{rcdj}

\begin{document}

\setcounter{page}{431}
\journal{REGULAR AND CHAOTIC DYNAMICS, V.\,8, \No4, 2003}
\runningtitle{GALTON BOARD}
\title{GALTON BOARD}
\runningauthor{V.\,V.\,KOZLOV, M.\,Yu.\,MITROFANOVA}
\authors{V.\,V.\,KOZLOV}
{Department of Mechanics and Mathematics\\
Moscow State University,
Vorob'ievy Gory\\
119899, Moscow, Russia}
\authors{M.\,Yu.\,MITROFANOVA}
{E-mail: mitrofanov@breitmeier.de}

\abstract{In this paper, we present results of simulations of a model of the
Galton board for various degrees of elasticity of the ball-to-nail collision.}
\amsmsc{37A50, 70F45, 70F35, 82C22}
\doi{10.1070/RD2003v008n04ABEH000255}
\received 1.10.2003.

\maketitle

\section{Introduction}

The Galton board is an upright board with evenly
spaced nails driven into its upper half. The
nails are arranged in staggered order. The lower half of the board is
divided with vertical slats into a number of narrow rectangular
slots. From the front, the whole installation is covered with a
glass cover. In the middle of the upper edge, there is a funnel in
which balls can be poured, the diameter of the balls
being much smaller than the distance between the nails. The funnel is
located precisely above the central nail of the second row, i.\,e. the
ball, if perfectly centered, would fall vertically and directly
onto the uppermost point of this nail's surface (Fig. 1).

\wfig<bb=0 0 50.8mm 47.9mm>{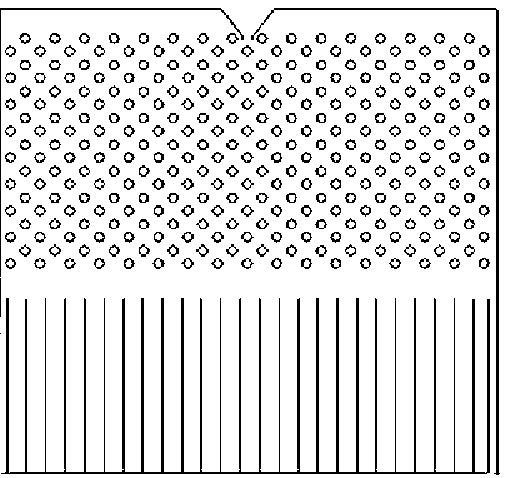}

Theoretically, the ball would repeatedly bounce off this nail's uppermost
point. Obviously, such a motion of the ball is unstable. In fact, due to
unavoidable inaccuracy in the board's positioning and impossibility to
completely exclude the lateral component, no matter how small, of the
ball's velocity, each ball, generally speaking, would meet the nail
somewhat obliquely. The ball would then deviate from the vertical line
and, after having collided with many other nails, fall into one of the
slots. If the experiment is run with a large number of balls,
dropped one after the other, then the following results are obtained: the
balls are distributed evenly to the left and to the right of the central
compartment (left and right deviations are equiprobable). Besides, the
balls would more rarely fall into the leftmost and rightmost compartments,
for large deviations are more rare to appear than small ones. However, despite  the presence of nails and all the imperfections in the construction, the majority of the balls
will agglomerate  in the central compartment as this provides
the smallest deviation. The number of  balls in the
compartments would approximately correspond to the Gaussian law of errors.
In the earlier experiments with the Galton board the funnel was
filled with pellets or millet grains.

\section{Statement of problem}

In Galton board experiments
ball-to-nail impacts have always been inelastic. In this paper, we present results of simulations of a model of the Galton board for
various degrees of elasticity of the ball-to-nail collision.

We model the ball as a mass point. Hence, the ball's motion can be
regarded as the motion of a mass point in a vertical plane under the
action of gravity accompanied  with multiple collisions with the nails. These
collisions are characterized by the coefficient of restitution~$e$, which
affects only the normal component of the ball's velocity after the impact.
The coefficient of restitution is the first parameter of the problem. It
can vary  from 0 to 1. A value of~$e=1$ corresponds to
absolutely elastic impact for which the ball's energy does not change. The
other extreme case,~$e=0$, corresponds to absolutely inelastic impact: the
ball ``sticks'' to the nail. The nail's radius~$R$ is the second parameter
of the problem. Since the ball leaves the funnel and falls onto the nail
centrally, but possibly with some small departure to the left or to the right, we adopt that the first drop of the ball  obeys the Gaussian law. On the other
hand, if  the balls distribute uniformly over the funnel's
opening, then their distribution over the rectangular compartments will be
far from normal (Fig.~2). This distribution resembles \emph{the arcsine
law\/}. Incidentally, according to Paul Levy, the distribution of time
intervals over which a Brownian particle is located  on the positive semi-axis, has a similar form. This observation is, probably, not just a
coincidence. The point is that a particle in Brownian motion experiences a
large number of random collisions with molecules of the surrounding fluid
(in our case, the ``molecules'' are regularly placed and fixed).

\fig{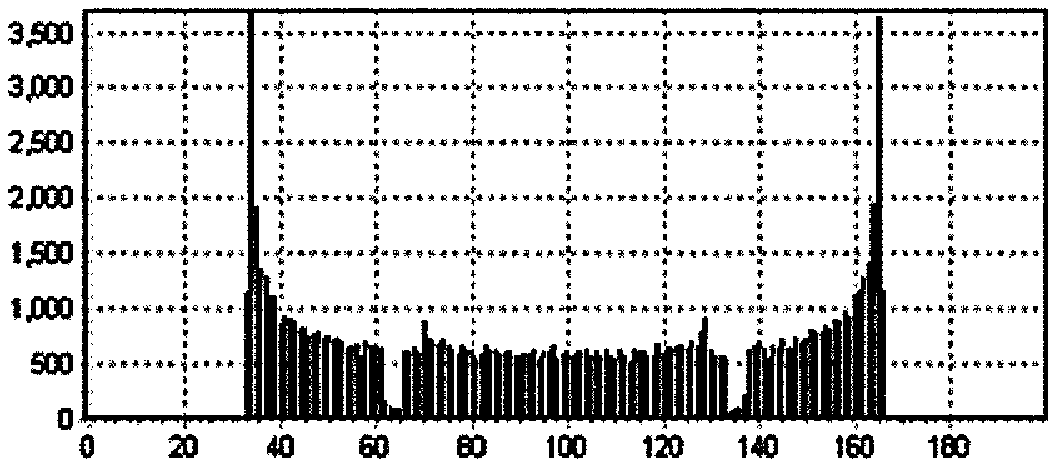}

Accordingly, the problem's third parameter is the variance of the
distribution of the  balls over the funnel's opening. Thus, we
introduce three parameters for the problem: 1) the coefficients of
restitution~$e$, 2) the nail's radius~$R$, 3) the variance~$\sigma_{0}$
of the normal distribution of the first ball-to-nail impact.

It is also necessary to choose the dimensions of the model board. The geometry of the board should meet the two requirements:
\begin{itemize}
  \item the balls should not reach the vertical boundaries of the Galton
  board;
  \item each ball should experience at least several collisions with the
  nails before it gets into one of the rectangular compartments.
\end{itemize}

\begin{figure}[!ht]
\begin{center}\it
\begin{tabular}{c c}
  \includegraphics[width=2.2in,height=1.1in]{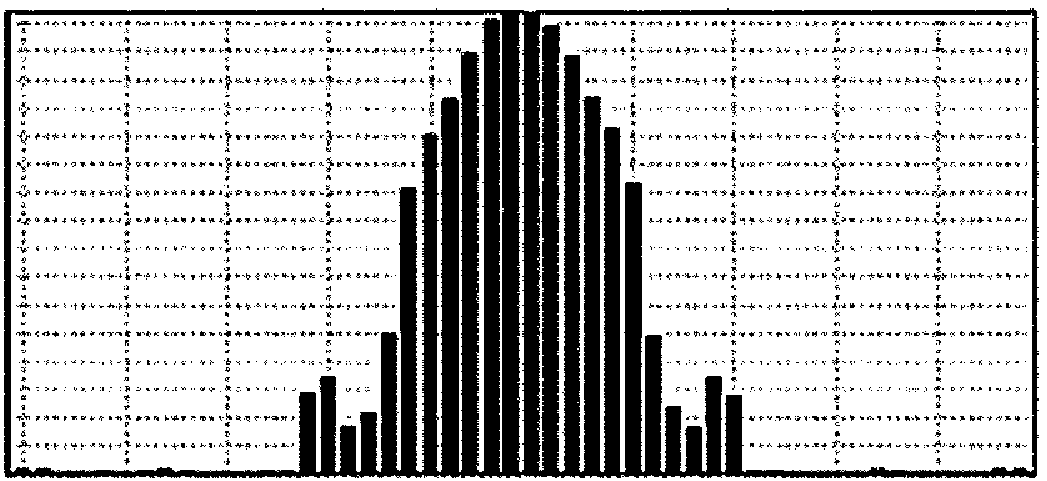} & \includegraphics[width=2.2in,height=1.1in]{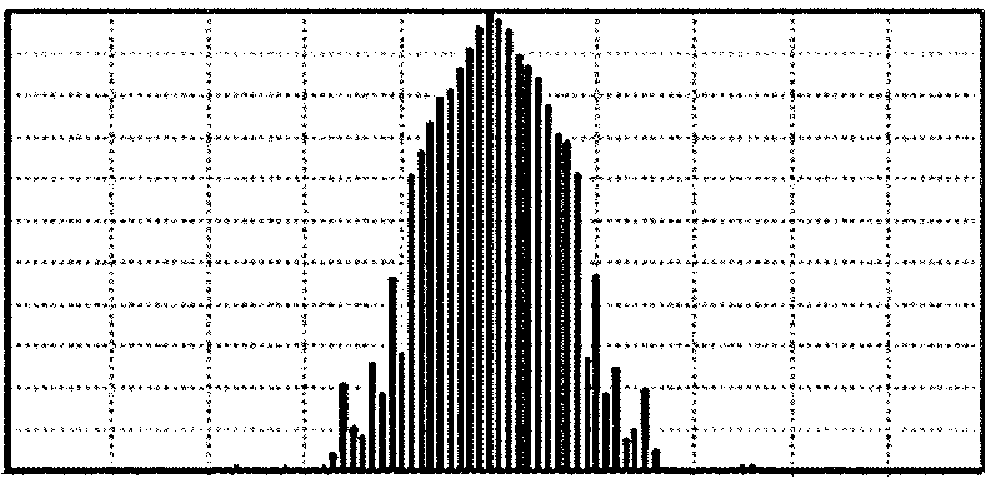} \\
  a & b \\
  \includegraphics[width=2.2in,height=1.1in]{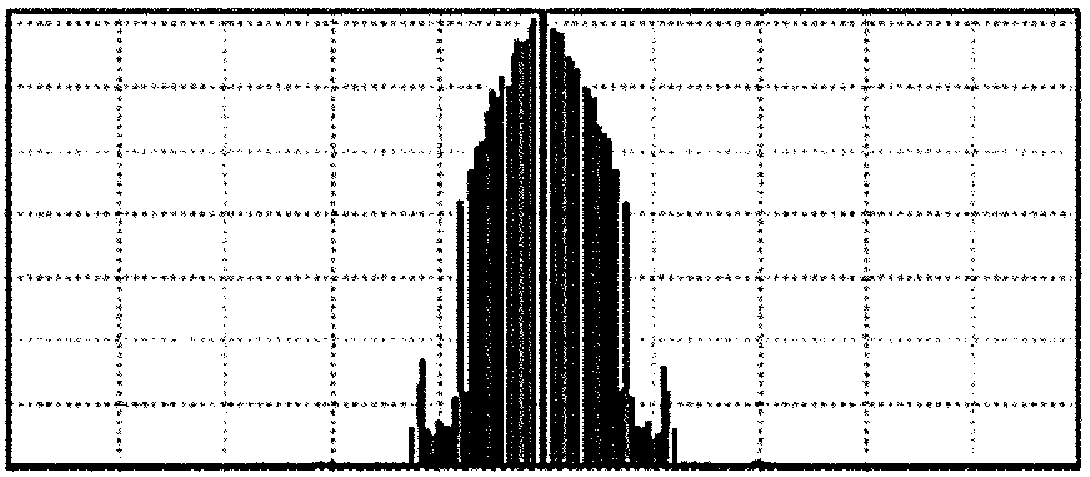} & \includegraphics[width=2.2in,height=1.1in]{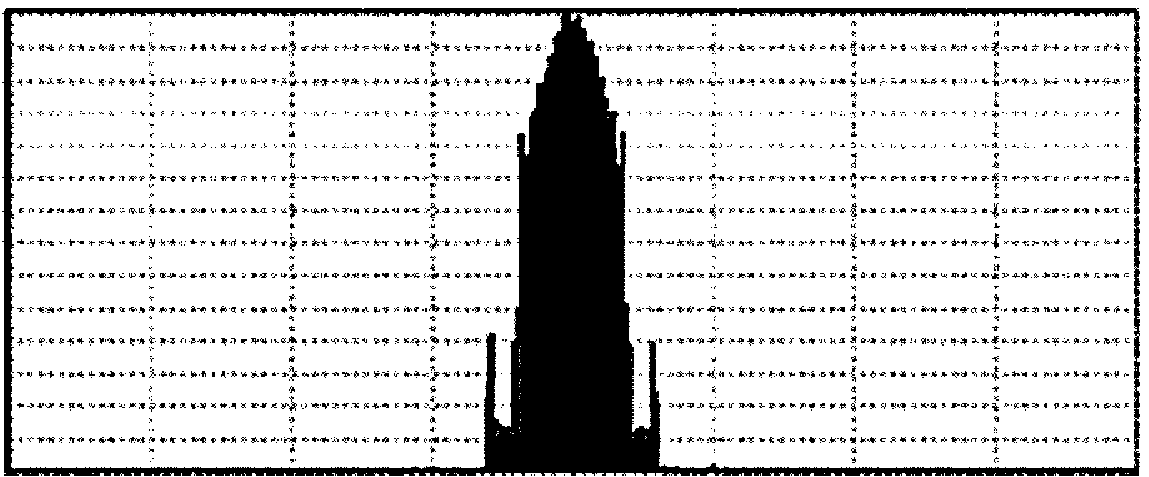} \\
  c & d \\
\end{tabular}
\end{center}\vspace{-3mm}
\caption{}
\end{figure}

Figure 3 (a--d) shows the balls' distribution over the
rectangular compartments for different dimensions of the model board,
namely,~$50 \times 50$ (a), $100 \times 100$ (b), $200 \times 200$ (c),
and $400 \times 300$ (d). In these cases, the three parameters of the
problem are:~$e=0.8$, $R=0.7$, and~$\sigma_{0}=0.05$. As we can see, the
distribution of the balls looks similarly for all the specified dimensions
of the board, but in the case of~$50 \times 50$ (Fig.~3a) the balls do reach the
vertical boundaries of the model Galton board. For the~$100 \times 100$
board, the balls no longer reach the boundaries (Fig.~3b). Further
enlargement of the model board's dimensions (its length and its
height) does not affect the pattern of the balls' distribution over the
compartments (Figs.~3c, d), but greatly increases the computation time.
Thus, the dimensions of the model Galton
board can be set to~$100 \times 100$ without any loss of quality.

So, we are going to investigate the properties of the balls' distribution
over the compartments of the Galton board and the dependency of the
variance of this distribution on the three specified parameters.

\section{Mathematical model}

The method of investigation consists in
simulating the motion of the balls (mass points) and taking into account
their collisions with obstacles (the nails) for different values of the
three specified parameters of the problem.

On the Galton board, we introduce an orthogonal coordinate system~$Oxy$ in
the following way: the axis~$Ox$ is directed horizontally and passes through
the upper boundaries of the rectangular compartments, in which the falling
balls are to be collected (for  brevity, from this point on, we
will say {\it compartments} instead of {\it rectangular compartments}).
The axis~$Oy$ is directed vertically and goes
through the center of the nail that a ball is to hit first. The length of
the board is taken large enough for the balls not to reach its vertical
boundaries (as was specified earlier).

Thus, a pair~$x,\,y$ represents the position of a ball in the plane of the
Galton board. The fall of the ball is described with a set of two ordinary
differential equations:
\begin{equation}\label{1}
  \ddot{x} = 0, \qquad    \ddot{y} = - g,
\end{equation}
where $g$ is the gravitational acceleration.

Since the ball falls from the funnel and onto the first nail under
gravity, the velocity of the ball at the point of the first impact is
~$v_0=\sqrt{2g(h_0-R\sin\alpha_0)}$, where~$h_0$ is the distance
between the funnel's opening and the center of the first nail,~$R$ is the
nail's radius, while~$\alpha_0$ is the angle between the axis~$Ox$ and the
radius drawn to the point where the ball hits the nail (Fig.~4).

\fig<bb=0 0 77.1mm 63.2mm>{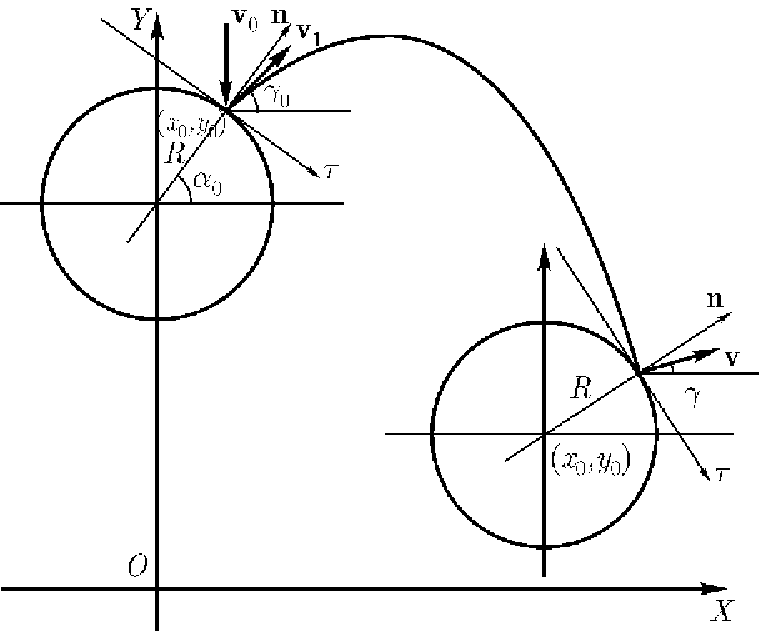}

We will investigate the further motion of the ball according to the
following plan:

\begin{enumerate}
  \item Introduce a coordinate system fixed to the nail: its origin is at
  the ball-to-nail impact point, and its axes are the tangent and the
  normal to the nail's surface at this point. Thus, with respect to this coordinate system, the velocity of the
  particle at the first impact point has the
  following components:~$v_0^{\tau}=v_0\cos\alpha_0$,
  $v_0^n=-v_0\sin\alpha_0$.
  \item After the impact with the nail, the velocity components will
  change and take the form: $v_1^{\tau}=v_0\cos\alpha_0$,
  $v_1^n=-ev_0^n=ev_0\sin\alpha_0$, where~$e$ is the coefficient of
  restitution.
  \item Then the ball will move in a parabola. To find its path, we solve
  the equations~(1) with the following initial values: $x\,(0)=x_0$,
  $y\,(0)=y_0$, $\dot{x}\,(0)=v_1\cos\gamma_0$,
  $\dot{y}\,(0)=v_1\sin\gamma_0$, where~$(x_0,\,y_0)$ are the coordinates
  of the ball at the time  it hits the
  nail,~$v_1=\sqrt{(v_1^{\tau})^2+(v_1^n)^2}$, while~$\gamma_0$ is the
  angle between the axis~$Ox$ and the velocity vector~$v_1$. Thus, the
  ball's path is the following parabola:
\begin{equation}\label{2}
  y = - \frac{g}{2v_{1}^{2}\cos^2\gamma_{0}}(x - x_{0})^{2} + (x -
  x_{0})\tan\gamma_{0} + y_{0}.
\end{equation}
  The portion of the parabola the ball will take is determined by the direction
  of the ball's velocity vector after its impact with the nail.
  \item From (1) we find the velocity  with which the ball will approach
  the next nail. Let~$(x_1,\,y_1)$ be the coordinates of the point of the
  next ball-to-nail impact. Then the velocity of the ball on the surface of
 this nail has the following components:
$$
  v_{2}^{x} = v_{1}\cos\gamma_{0}, \qquad v_{2}^{y} = - \frac{g(x_{1} -
  x_{0})}{v_{1}\cos\gamma_{0}}+ v_{1}\sin\gamma_{0}.
$$
\end{enumerate}
Then another collision occurs, and again the ball's motion is calculated
according to the procedure described above. The whole operation is
repeated until the ball crosses the axis~$Ox$. As soon as the ball's path
crosses the axis~$Ox$, we find the intersection point and thus determine the
compartment the ball  falls into.

One of the most important things about this model is to find the nail that
the ball is going to hit next. To that end, consider
 the perpendiculars to the ball's path which  go through the nails'
centers. Such perpendiculars are described with linear equations:
\begin{equation}\label{3}
  x - x_{n} + \left( - \frac{g}{v^{2}\cos^{2}\gamma}(x_{n}-x_{\text{imp}})+
  \tan\gamma\right)(y - y_{n}) = 0,
\end{equation}
where $(x_{\text{imp}},\,y_{\text{imp}})$ are the coordinates of the
previous ball-to-nail impact, $(x_n,\,y_n)$ are the coordinates of the
path's point through which a perpendicular is drawn, $v$ is the magnitude of the
ball's velocity after the previous impact, and $\gamma$ is the angle
between the axis~$Ox$ and the velocity vector~$v$.

Since our goal is to find  perpendiculars  through the nails'
centers, we insert the coordinates of the center of one of the
nails~$(x_{c},\,y_{c})$ into~(3). From this equation, we find a
pair~$(x_{n},\,y_{n})$ which meets the following requirements:
\begin{itemize}
  \item the distance between the nail's center and the
  point~$(x_{n},\,y_{n})$ is smaller than the nail's radius;
  \item the absolute value of the difference between the abscissa of the
  previous impact point and the abscissa of the path's point, through which the
  perpendicular is drawn, is as small as possible.
\end{itemize}

The first requirement is to ensure that the ball's path meets the nail,
i.\,e.  the coordinates of the next impact
point~$(x,\,y)$ can be found. These coordinates satisfy the following set of
equations:
$$
\label{5}  \left\{
\begin{array}{cc}
(x - x_{c})^{2} + (y - y_{c})^{2}  =  R^{2},  \\
y  =  - \frac{g}{2v^{2}\cos^{2}\gamma}(x - x_{\text{imp}})^{2} + (x - x_{\text{imp}})\tan\gamma + y_{\text{imp}},  \\
\end{array}
\right.
$$
where $(x_{c},\,y_{c})$ are the coordinates of the nail's center, $R$ is
the nail's radius, and~$(x_{\text{imp}},\,y_{\text{imp}})$ are the
coordinates of the previous impact point.

The second requirement is to take the impacts in their sequence. This
follows from the parametric form of the ball's path. We solve the
equations~(1) with the following initial values:~$x\,(0)=x_{\text{imp}}$,
$y\,(0)=y_{\text{imp}}$, $\dot{x}\,(0)=v\cos\gamma$,
$\dot{y}\,(0)=v\sin\gamma$, where~$(x_{\text{imp}},\,y_{\text{imp}})$ are
the coordinates of the previous impact point,~$v$ is the magnitude  of the
ball's velocity after the previous impact, while~$\gamma$ is the angle
between the axis~$Ox$ and the velocity vector~$v$. The result is the
parametric form of the ball's path after it has hit the nail:
$$
x = v t \cos\gamma + x_{\text{imp}}, \qquad y = - \frac{g}{2} t^2 + v
t\sin\gamma + y_{\text{imp}}.
$$

\textbf{5.  Simulation results.} The model Galton board has been
implemented as an interactive Microsoft Visual
C++ application. The application offers the opportunity to vary every parameter of the
model: the nail's radius, the coefficient of restitution, and the
variance of the initial distribution when the ball hits the first nail;
it also allows varying the number of dropped balls. The application
outputs a histogram of the  balls in the compartments and the  variance  of the final
distribution of the balls over the compartments. Besides, the experiment's
results can be visualized.

First, we get a histogram of the distribution of the balls over the
compartments (Fig.~5). Form this histogram, the variance is calculated
using the following well-known formulas:
$$
\bar{x} = \sum\limits_{i=1}^n x_{i} \frac{N_{i}}{N}, \qquad
\sigma^{2} = \sum\limits_{i=1}^{n} (x_{i} - \bar{x})^{2}
\frac{N_{i}}{N},
$$
where $\bar{x}$ is the mathematical expectation, $x_{i}$ is the coordinate
of the center of the~\textit{i}th compartment,~$n$ is the number of
compartments,~$N$ is the number of dropped balls,~$N_{i}$ is the final number of
balls accumulated in the~\textit{i}th compartment, and~$\sigma$ is the
variance of the distribution.

Second, using the variance obtained, we plot the normal distribution
(Gaussian) curve using the  Gauss
formula~$f\,(x)=\frac1{\sigma\sqrt{2\pi}}e^{-(x-\bar{x})^2/(2\sigma^2)}$.
Next, we compare the theoretical curve with the model curve (Fig. 6). The
model curve is plotted with squares, while the theoretical curve is
plotted as a solid line.

\ffig<width=0.48\textwidth>{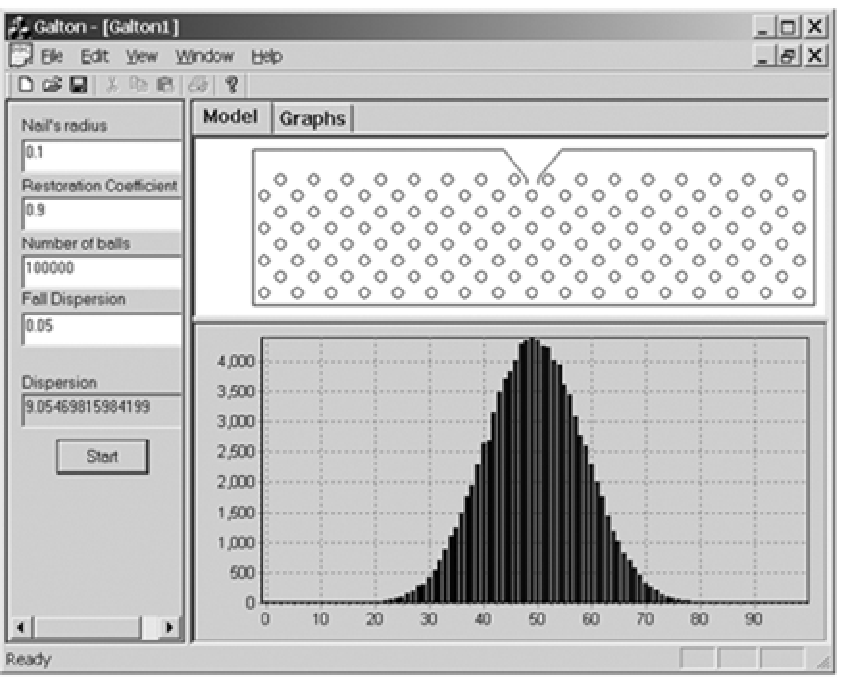}<width=0.48\textwidth>{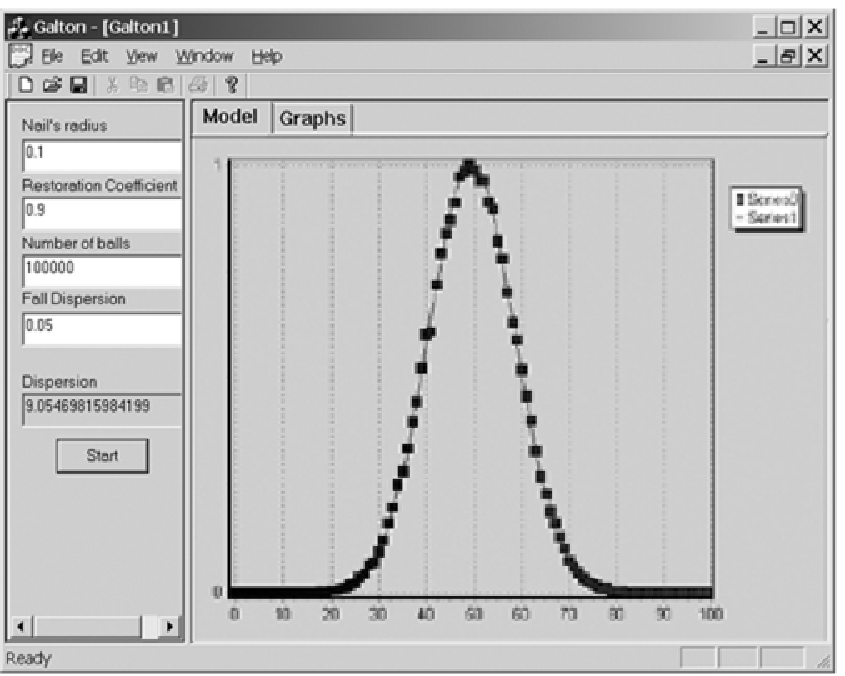}

The results given in this paper were obtained for~100\,000 dropped balls.
This number is optimal from the viewpoint of the result's accuracy and the
processing time needed for the experiment. Figures~7 (a--d) show the
histograms of the balls' distribution over the compartments for~1\,000~(a),
10\,000~(b), 100\,000~(c), and~1\,000\,000 (d) of dropped balls.
These results were obtained with the following value of the parameters:~$e=1$, $R=0.1$, and~$\sigma_0=0.04$. We can see that the
histograms shown in Figs.~7c and d are, for all practical purposes,
identical. The accuracy of the  results can also be judged
by the figures given in Table~1. These figures are the values of the
variance of the final distribution for~10\,000, 100\,000 and~1\,000\,000
dropped balls in eight series of simulations.

\begin{figure}[!ht]
\begin{center}\it
\begin{tabular}{c c}
  \includegraphics[width=2.2in,height=1.1in]{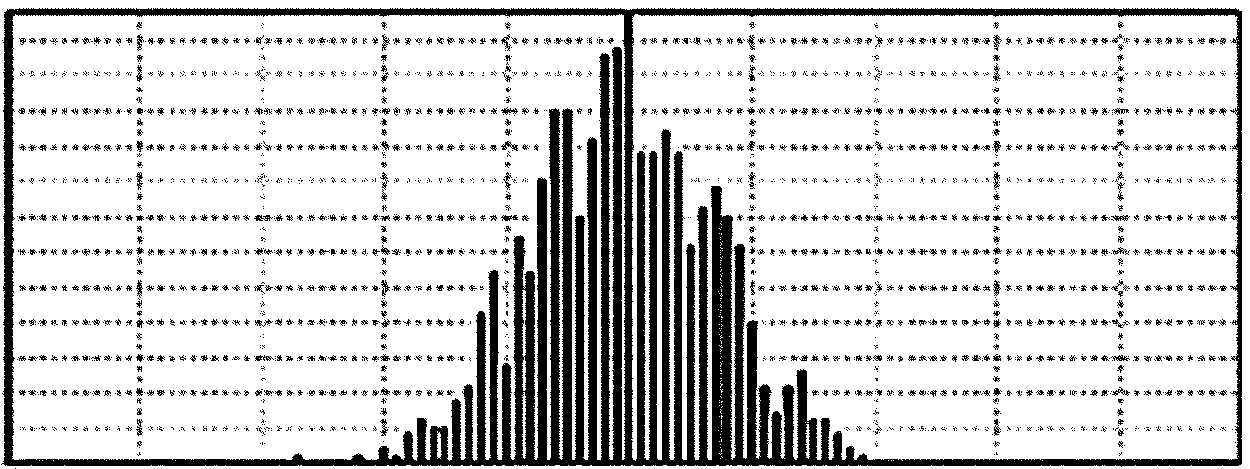} & \includegraphics[width=2.2in,height=1.1in]{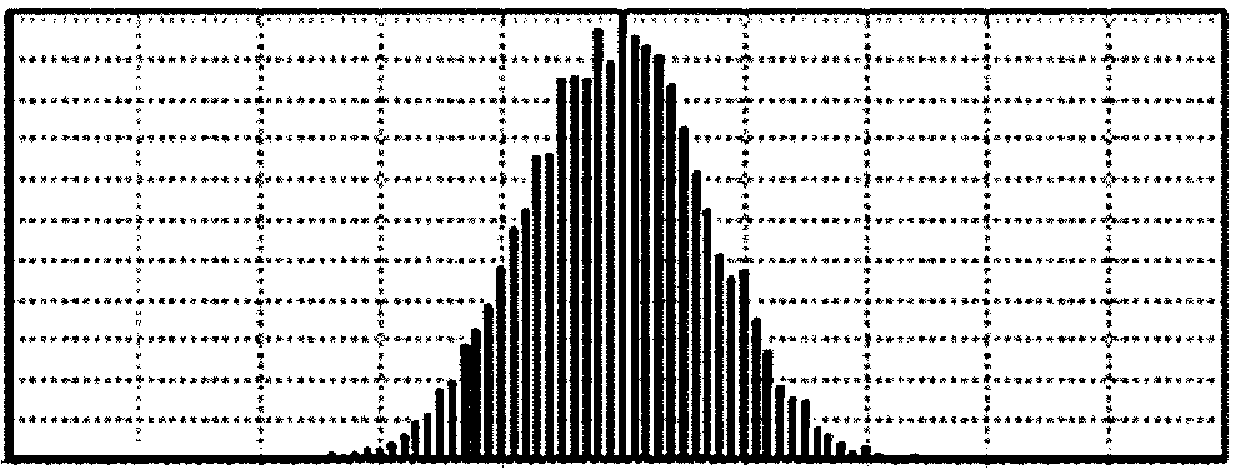} \\
  a & b \\
  \includegraphics[width=2.2in,height=1.1in]{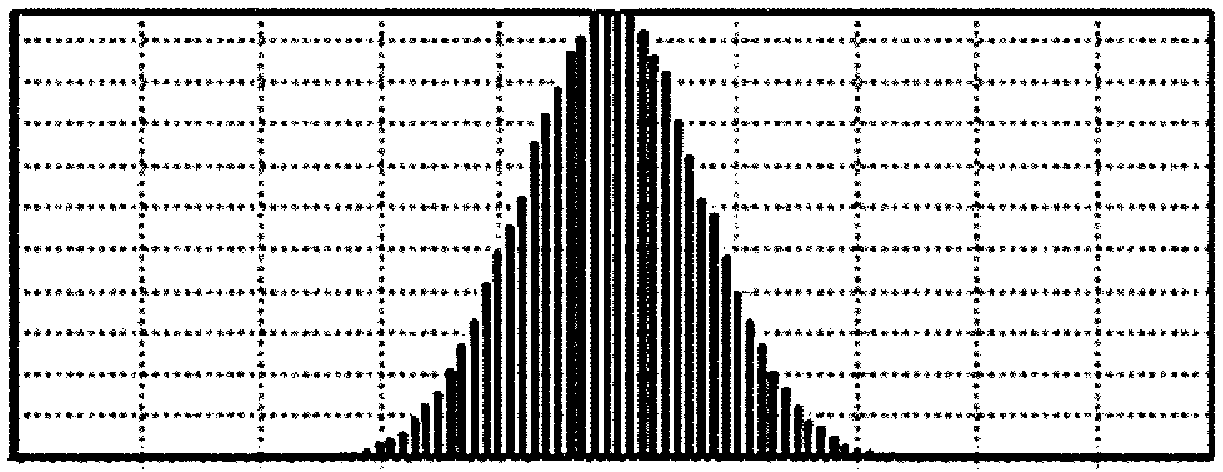} &
  \includegraphics[width=2.2in,height=1.1in]{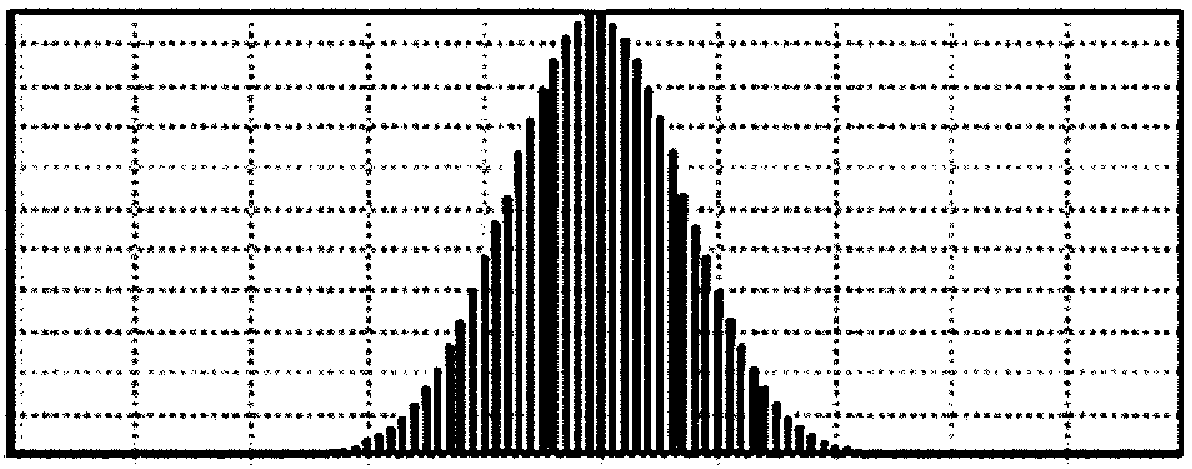} \\
  c & d \\
\end{tabular}
\end{center}\vspace{-3mm}
\caption{}\vspace{-6mm}
\end{figure}

\begin{table}[!ht]
\caption{$e = 1$, $R = 0.1$, and $\sigma_{0} = 0.04$}
\begin{center}
\begin{tabular}{|c|c|c|c|c|c|c|c|c|}
  \hline
  N       & 1      & 2      & 3      & 4      & 5      & 6      & 7      & 8      \\ \hline
  10000   & 7.3820 & 7.3381 & 7.3367 & 7.4097 & 7.4266 & 7.4201 & 7.3159 & 7.3559  \\ \hline
  100000  & 7.4066 & 7.3817 & 7.3841 & 7.3938 & 7.3779 & 7.4387 & 7.4063 & 7.3713  \\ \hline
  1000000 & 7.4028 & 7.3988 & 7.3901 & 7.3917 & 7.4066 & 7.3958 & 7.3902 & 7.4104  \\ \hline
\end{tabular}
\vspace{-3mm}
\end{center}
\end{table}

Let us first consider the case where the balls are distributed uniformly on
 the width of the funnel's opening. Instead of the  normal
distribution of the balls over the compartments (as it might be expected), we get a somewhat unusual distribution with peripheral peaks and
two distinctive gaps near the center (Fig.~2). These gaps are located symmetrically with respect to the vertical axis  through the funnel's center. For this
case, the~$200 \times 100$ model Galton board was taken, otherwise the
balls reach its vertical boundaries.

More precisely, Fig.~2 corresponds to the case of absolutely elastic
impact ($e=1$) and~$R=0.1$.

\begin{figure}[!ht]
\begin{center}\it
\begin{tabular}{c c}
  \includegraphics[width=2.2in,height=1.1in]{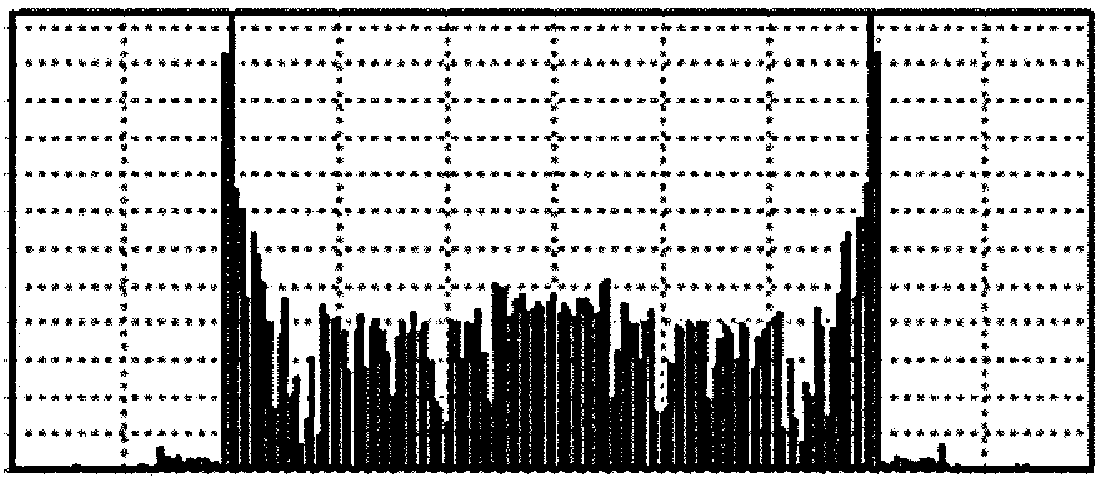} & \includegraphics[width=2.2in,height=1.1in]{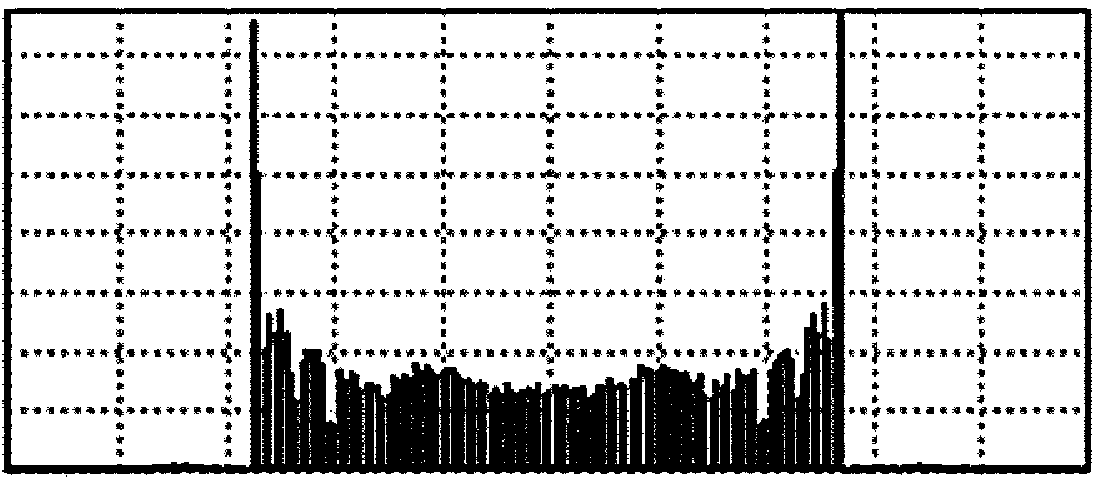} \\
  a & b \\
  \includegraphics[width=2.2in,height=1.1in]{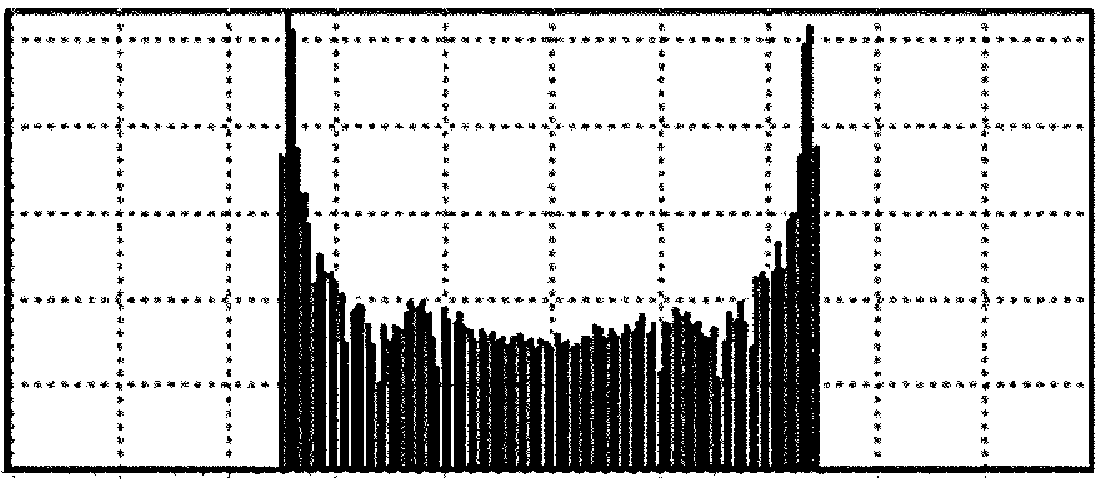} & \includegraphics[width=2.2in,height=1.1in]{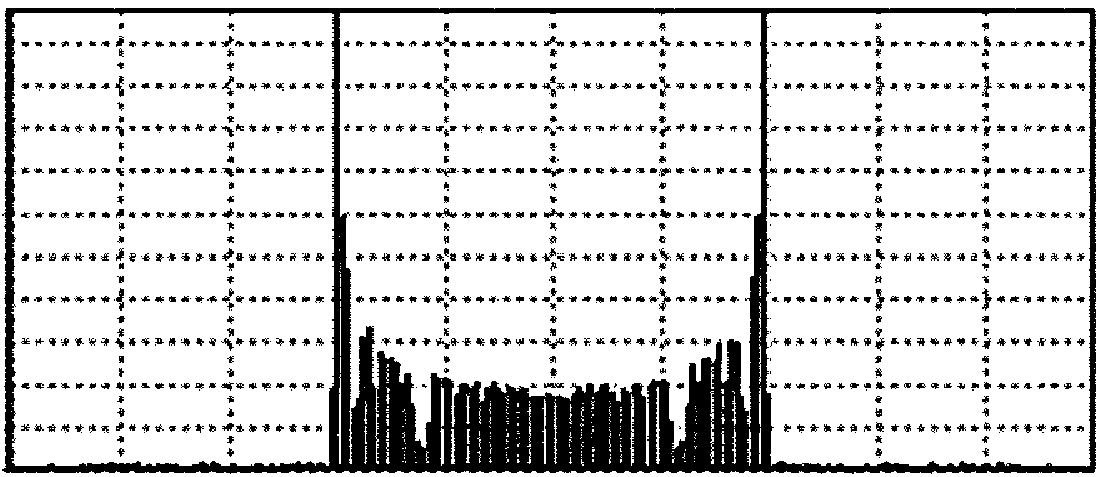} \\
  c & d \\
\end{tabular}
\end{center}\vspace{-3mm}
\caption{}
\end{figure}

For a smaller value of  the coefficient of restitution ($e=0.8$) and an increased value of the
nail's radius to~$R=0.3$,  the balls' distribution over the
compartments   changes not very significantly:
distinctive peripheral  peaks are still present, while instead of two pronounced gaps we
have several symmetrically located small pits (Fig.~8a). With a further
decrease of the coefficient of restitution the depth and structure of
these pits changes. In Figs.~8b, c, and d, the histograms are shown
for~$e=0.6$, $e=0.4$, and~$e=0.1$, respectively (the nail's radius is the same,~$R=0.3$).

If the balls are fed into the  funnel according to a  Gaussian law with large
dispersion~$\sigma_0$, then the form of the histograms will not change
qualitatively. Therefore, the case of small dispersion~$\sigma_0$ becomes
especially interesting.

Let~$\sigma_0=0.05$ and~$R=0.4$. We are going to investigate the
form of the histogram, as  the coefficient of restitution~$e$ decreases
from 1 to 0. Figures~9 (a--d) show the balls' distributions over the
compartments for~$e=1$~1 (a), 0.9 (b), 0.8 (c), and~0.7 (d).

\begin{figure}[!ht]
\begin{center}\it
\begin{tabular}{c c}
  \includegraphics[width=2.2in,height=1.1in]{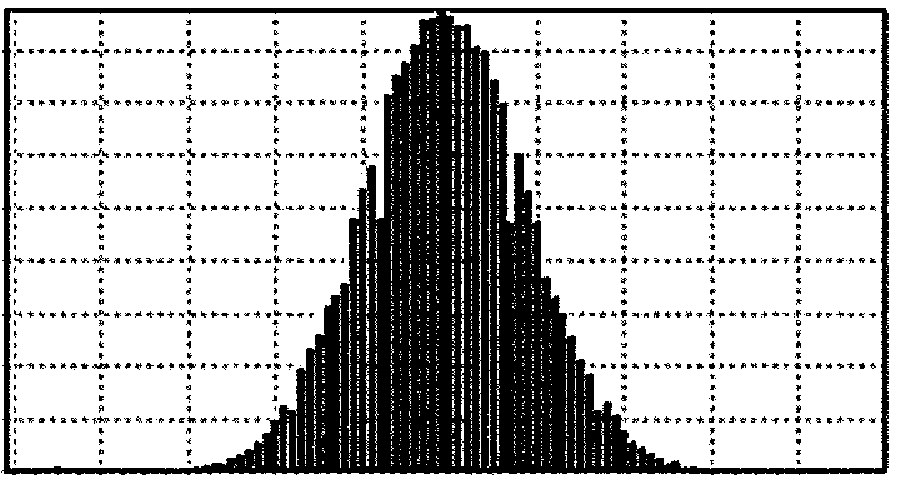} & \includegraphics[width=2.2in,height=1.1in]{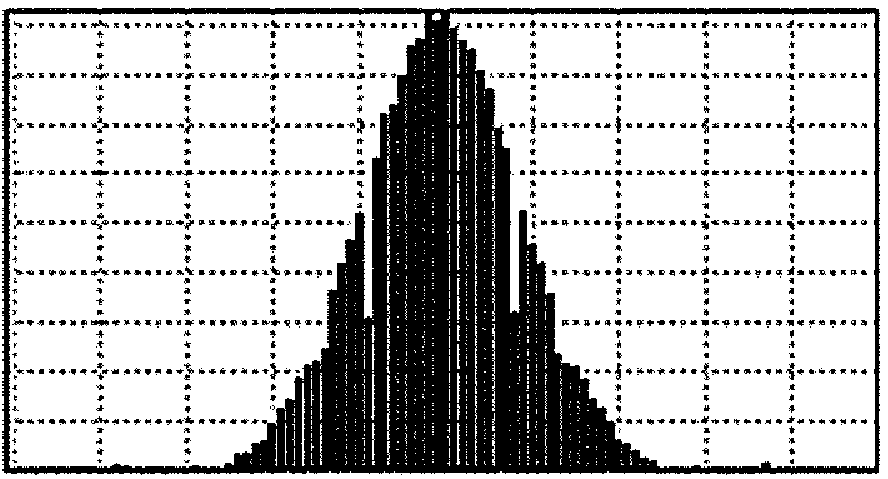} \\
  a & b \\
  \includegraphics[width=2.2in,height=1.1in]{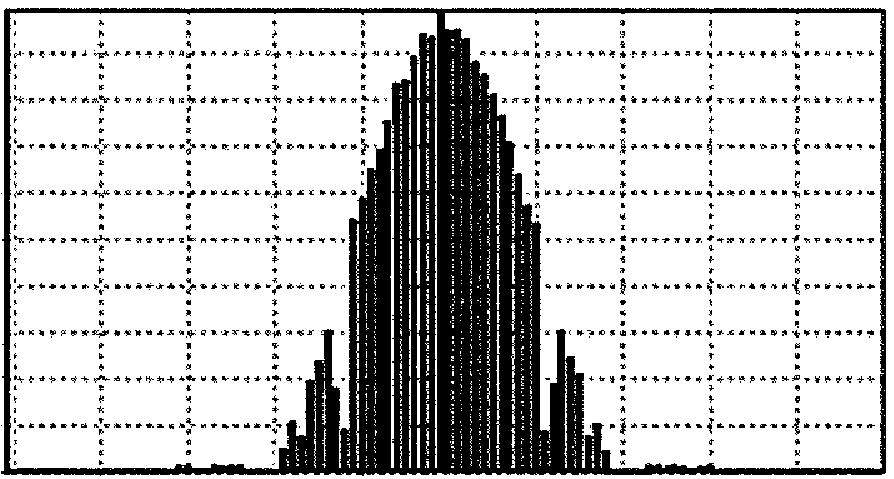} & \includegraphics[width=2.2in,height=1.1in]{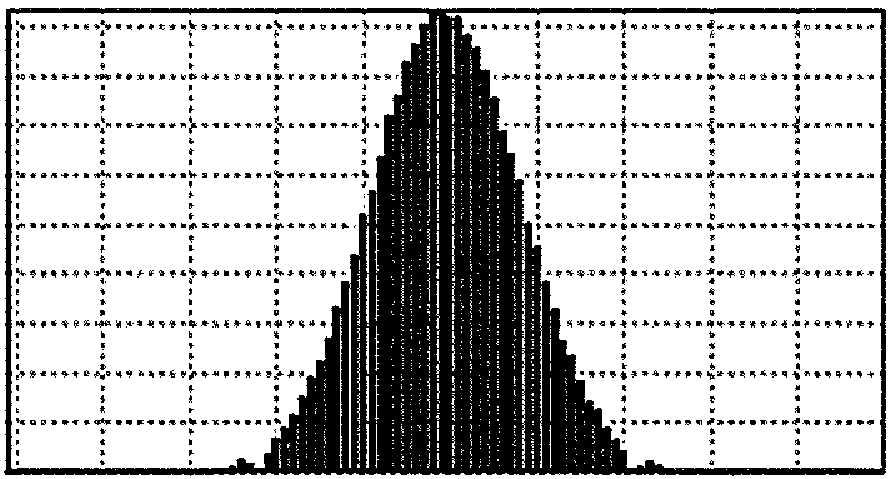} \\
  c & d \\
  \multicolumn{2}{c}{\includegraphics[width=2.2in,height=1.2in]{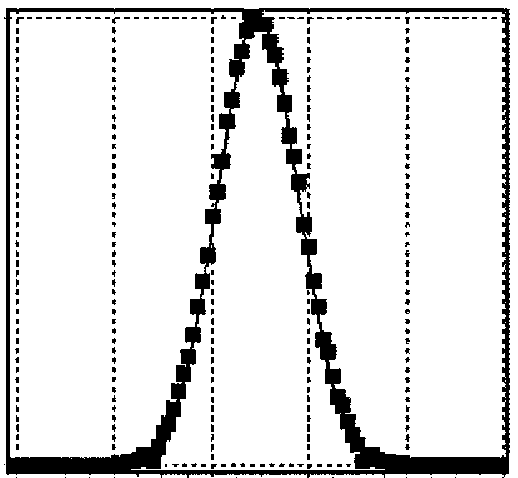}} \\
  \multicolumn{2}{c}{e}
\end{tabular}\vspace{-4mm}
\end{center}
\caption{}\vspace{4mm}
\end{figure}

We can see that the distribution in the case of absolutely elastic impact
is almost Gaussian  with two noticeable pits. As the coefficient of
restitution decreases, the ``normal'' distribution gets ``corrupted'';
instead of the pits, distinctive gaps appear, which become deeper with a
decrease of~$e$. However, this picture holds only for $e$ below a value
of~$e\approx 0.7$. A further decrease of  the coefficient of restitution
makes the gaps disappear, and the distribution becomes practically
indistinguishable from the Gaussian distribution (Fig.~9e).

Figures~10 (a--e) show a similar series of histograms for~$R=1$, while the coefficient of restitution takes  successively
the values~0.7, 0.6, 0.3, 0.2, and~0.1. We can see that for~$e=0.2$ there
are two gaps, while for  larger and smaller values
of~$e$ the distribution is, practically, Gaussian. A further increase in~$e$
results in a distribution which is very different from Gaussian.

\begin{figure}[!ht]
\begin{center}\it
\begin{tabular}{c c}
  \includegraphics[width=2.2in,height=1.1in]{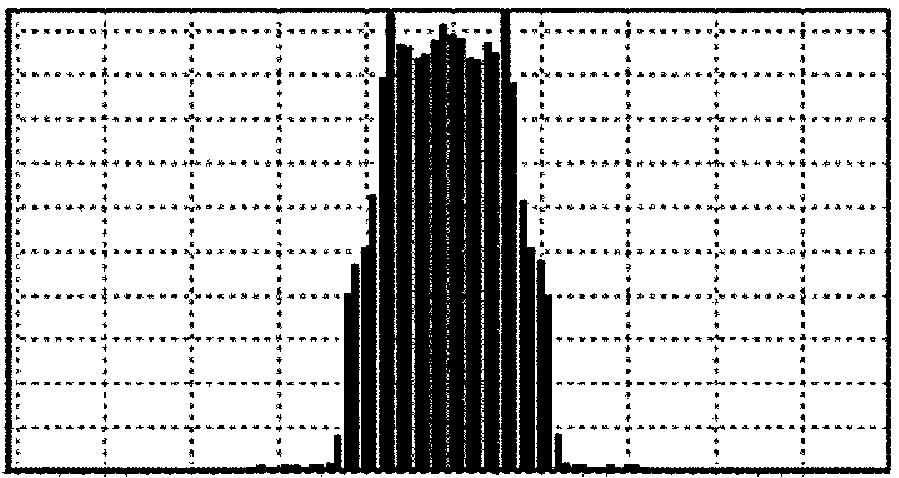} & \includegraphics[width=2.2in,height=1.1in]{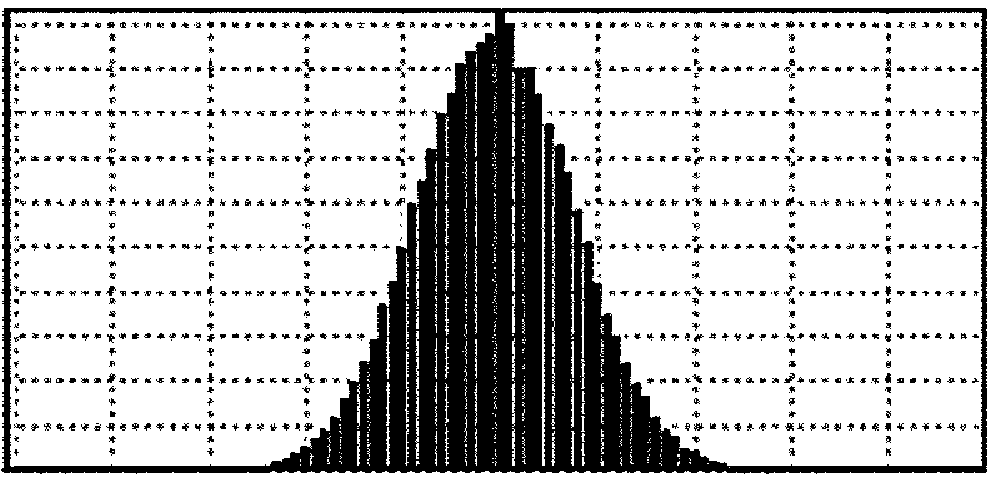} \\
  a & b \\
  \includegraphics[width=2.2in,height=1.1in]{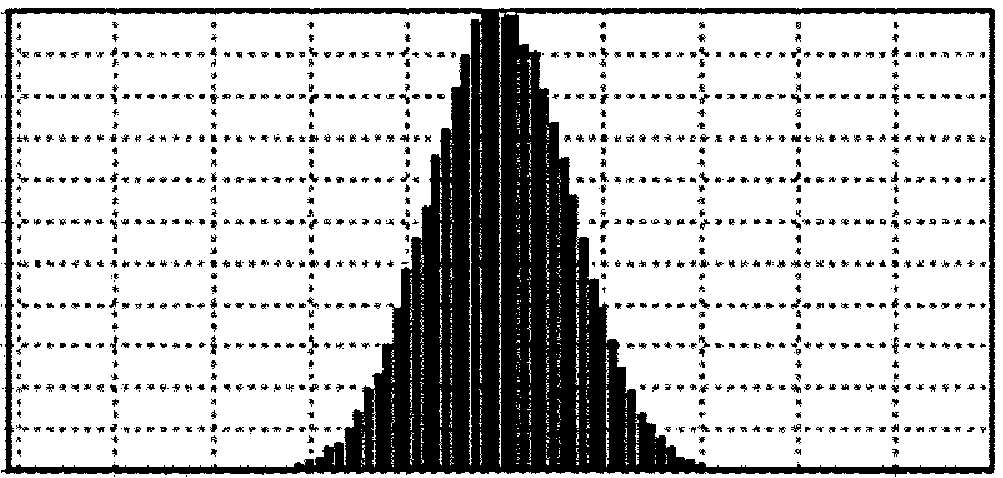} & \includegraphics[width=2.2in,height=1.1in]{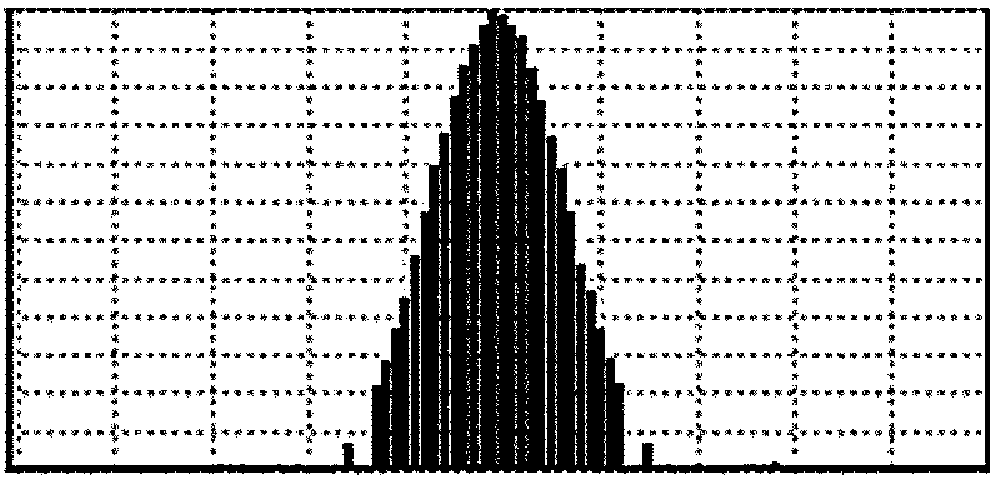} \\
  c & d \\
  \multicolumn{2}{c}{\includegraphics[width=2.2in,height=1.2in]{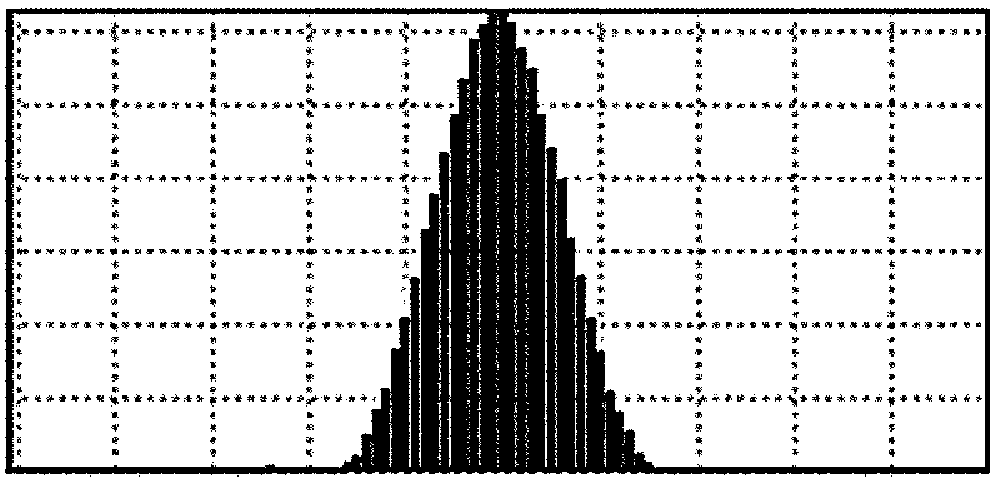}} \\
  \multicolumn{2}{c}{e}
\end{tabular}
\end{center}\vspace{-3mm}
\caption{}
\end{figure}

The mentioned ``occasions'' of deviation of the final distribution of the
balls from a Gaussian distribution are intriguingly associated with non-monotonic behavior
of the variance~$\sigma$  as a function of
two variables~$R$ and~$e$ (with~$\sigma_0$ being fixed). Table~2 gives the
values of~$\sigma$ for~$\sigma_0=0.05$. We can see that for a fixed~$R$ ,
the variance ~$\sigma$ a local maximum. It is exactly that value of the coefficient of restitution in the vicinity of which  a
substantial deviation from the Gaussian distribution occurs. For example,
for~$R=0.4$, the variance~$\sigma$ has a maximum at~$e\simeq 0.7$; this
value has already been mentioned in connection with the analysis of the
series of histograms in Fig.~9. Similarly, for~$R=1$, the local maximum is
reached at~$e\simeq 0.2$ (as it  should be, according to
Fig.~10).

\begin{table}[!ht]
\caption{$\sigma_0=0.05$}
\begin{center}
\begin{tabular}{|c|c|c|c|c|c|c|c|c|c|c|}
\hline
  $e \setminus R$ & 0.1 & 0.2 & 0.3 & 0.4 & 0.5 & 0.6 & 0.7 & 1 & 1.2 & 1.5 \\
  \hline
    1 & 9.24 & 9.19 & 9.26 & 9.44 & 9.61 & 9.94 & 10.29 & 10.38 & 11.27 & 12.80 \\ \hline
  0.9 & 9.06 & 8.97 & 9.05 & 9.18 & 9.22 & 9.32 & 9.57 & 10.63 & 12.27 & 11.36 \\ \hline
  0.8 & 8.79 & 8.47 & 8.32 & 8.07 & 7.83 & 7.52 & 7.78 & 8.22 & 8.05 & 8.47 \\ \hline
  0.7 & 8.57 & 8.56 & 8.46 & 8.41 & 8.34 & 8.26 & 8.22 & 7.14 & 7.69 & 8.82 \\ \hline
  0.6 & 8.33 & 8.26 & 8.30 & 8.21 & 8.18 & 8.15 & 8.10 & 7.96 & 7.89 & 7.72 \\ \hline
  0.5 & 8.06 & 8.00 & 7.98 & 7.94 & 7.91 & 7.84 & 7.79 & 7.65 & 7.60 & 7.75 \\ \hline
  0.4 & 7.77 & 7.72 & 7.68 & 7.61 & 7.64 & 7.57 & 7.52 & 7.36 & 7.22 & 7.06 \\ \hline
  0.3 & 7.53 & 7.42 & 7.34 & 7.30 & 7.22 & 7.23 & 7.17 & 7.02 & 6.86 & 6.76 \\ \hline
  0.2 & 7.03 & 7.08 & 7.05 & 7.07 & 7.08 & 7.08 & 7.18 & 7.23 & 7.20 & 6.64 \\ \hline
  0.1 & 6.67 & 6.72 & 6.88 & 7.03 & 6.94 & 6.96 & 6.83 & 6.68 & 6.68 & 6.72 \\ \hline
\end{tabular}
\end{center}
\end{table}

The specified features of the histograms require  theoretical
treatment and interpretation. The problem of the gaps should be especially
emphasized because this problem is likely to be most directly relevant to
the famous Kirkwood gaps in the distribution of asteroids in the main
asteroid belt between Mars and Jupiter. It is well known that these gaps
cannot be satisfactorily explained with the resonance ratios of the
orbital periods of the major planets. Meanwhile it would be useful to
investigate a simple model, where  small planets (large asteroids) move
along circular orbits, and there also is a flow of small asteroids
 colliding with the large ones without perturbing their paths. In this case, the
impact is not absolutely elastic ($0<e<1$). After a large
number of collisions, one would obtain a distribution of the asteroids'
flow over the semi-major axes. This distribution may contain a series of
gaps, as that in  the case of Galton board.

This work was partially supported by the grant ``Principal Scientific
Schools'' (00-15-96146).

\end{document}